# Dislocation pipe diffusion and solute segregation during the growth of metastable GeSn

Jérôme Nicolas,[1] Simone Assali,[1] Samik Mukherjee,[1] Andriy Lotnyk,[2] Oussama Moutanabbir[1,*]

[1] Department of Engineering Physics, École Polytechnique de Montréal, C. P. 6079, Succ. Centre-Ville, Montréal, Québec H3C 3A7, Canada
[2] Leibniz Institute of Surface Engineering (IOM), Permoserstr. 15, 04318 Leipzig, Germany

**Abstract.**

Controlling the growth kinetics from the vapor phase has been a powerful paradigm enabling a variety of metastable epitaxial semiconductors such as Sn-containing group IV semiconductors (Si)GeSn. In addition to its importance for emerging photonic and optoelectronic applications, this class of materials is also a rich platform to highlight the interplay between kinetics and thermodynamic driving forces during growth of strained, nonequilibrium alloys. Indeed, these alloys are inherently strained and supersaturated in Sn and thus can suffer instabilities that are still to be fully elucidated. In this vein, in this work the atomic-scale microstructure of $Ge_{0.82}Sn_{0.18}$ is investigated at the onset of phase separation as the epitaxial growth aborts. In addition to the expected accumulation of Sn on the surface leading to Sn-rich droplets and sub-surface regions with the anticipated equilibrium Sn composition of 1.0at.%, the diffusion of Sn atoms also yields conspicuous Sn-decorated filaments with nonuniform Sn content in the range of ~1 to 11at.% . The latter are attributed to the formation and propagation of dislocations, facilitating the Sn transport toward the surface through pipe diffusion. Furthermore, the interface between the Sn droplet and GeSn shows a distinct, defective layer with Sn content of ~22at.%. This layer is likely formed by the expelled excess equilibrium Ge as the Sn solidifies, and its content seems to be a consequence of strain minimization between tetragonal Sn-rich and cubic Ge-rich equilibrium phases. The elucidation of these phenomena is crucial to understand the stability of GeSn semiconductors and control their epitaxial growth at a uniform composition.

# INTRODUCTION

Sn-containing group IV semiconductors are an emerging class of materials for the monolithic integration of optoelectronic devices on silicon. Recently, major efforts have been expended to optimize their structural and optical properties.[1–4] It is now established that one of the main hurdles in their epitaxial growth has to do with the limited solubility of Sn in SiGe (<~1at.%), which is about an order of magnitude smaller than the Sn critical content to reach a direct bandgap in GeSn.[5,6] This is partially attributed to the large difference (~14.7%) in covalent radius between Ge (1.225Å) and Sn (1.405Å), which also causes a large compressive strain in GeSn when grown on Ge or Si.[7,8] The growth of supersaturated GeSn incorporating more than 1at.% of Sn is achieved using far-from-equilibrium growth conditions,[9–12] which prevent the formation of a biphasic mixture of Ge-rich and Sn-rich compounds in the early growth stage.[13,14] Yet, as the layers grow thicker, the alloy can undergo phase separation, leading to the nucleation of surface Sn droplets and material degradation.[15–17] It was shown that strain-relaxed buffer layers can delay surface segregation and phase separation in relatively thicker GeSn layers.[16] However, the exact mechanisms are yet to be elucidated especially in presence of partial plastic relaxation. Understanding the key atomic pathways of these phenomena is not only crucial to the optimization of material quality and device processing, but it is also of paramount importance to establish the structural stability of this material system that is inherently metastable and strained. In this vein, this work uncovers atomistic-level details at the onset of phase separation during the growth of metastable GeSn and addresses the associated mechanisms and structural properties. The role of dislocations in the instability of GeSn composition is identified, their interaction with solute atoms and their contribution to phase separation are unraveled and discussed.

**EXPERIMENTAL SECTION**

The GeSn layers investigated in this work were grown on Si(100) wafers in a low-pressure chemical vapor deposition (CVD) reactor using ultra-pure $H_2$ carrier gas, with monogermane diluted at 10% in $H_2$ and tin tetrachloride as precursors. First, a ~650 nm-thick Ge layer was grown on the Si wafer and subjected to thermal cycling annealing to improve its crystalline quality and surface roughness. Then, on top of this Ge virtual substrate (Ge VS), three successive GeSn layers were grown: the bottom layer (BL), the middle layer (ML) and the top layer (TL) at 320, 300 and 280°C, respectively.[18] The thickness of different layers was estimated using energy-dispersive X-ray spectroscopy (EDS) in a transmission electron microscope (TEM), which also provides detailed information on the microstructure. Composition and strain values for all layers were estimated from high-resolution X-Ray Diffraction (XRD) Reciprocal Space Mapping (RSM), using a bowing parameter $b = 0.041$ Å. Detailed atomic-level studies of the grown GeSn layers were carried out using laser-assisted atom probe tomography (APT). The sample preparation for APT was performed in Helios Nanolab 650 dual channel ($Ga^+$ ion column for milling, SEM column for imagining) focused-ion beam (Dual-FIB) microscope, using the standard lamella lift-out technique. Prior to the APT tip fabrication in Dual-FIB, a 50 nm thick Ni capping layer was co-deposited on all the samples (using an electron-beam evaporator) in order to protect the top-most part of the samples from ion-implanted damage during the tip fabrication process. APT achieves electric field-induced evaporation of atoms as cations, in a layer-by-layer fashion, from the surface of a needle-like specimen, with the assistance of an ultra-fast pulsed laser. In this work, the field evaporation of individual atoms in the APT was assisted by focusing a picosecond pulsed UV laser ($\lambda = 355$nm), with a beam waist smaller than 5µm, on the apex of the needle-shaped specimen. The laser pulse repetition-rate was maintained at 500 kHz throughout. The evaporation rate (ion/pulse) and the pulse energy were varied over a single run. An APT run started with the onset of evaporation of Ni atoms from the capping layer. During this, an evaporation rate of 0.8-1.0 and a laser pulse energy of 30.0 pJ was maintained. As soon as the atoms from the SL appeared at the outer rim of the detector ion map, the evaporation rate was slowed down to 0.2 in a single step and the laser energy was lowered to 4.0-5.0 pJ, in steps of 1.0 pJ. The run was slowed down to ensure that the tip makes a gradual and smooth transition from the Ni cap into the SL without fracturing it. When all the Ni atoms were evaporated and the transition into the SL was complete, the evaporation rate was slowly increased in small steps of 0.20 up to 1.0, ensuring in each step that the automatic voltage ramp is not too steep, a scenario which is well-known to cause tip fracture in APT. After the evaporation made a complete transition from the SL into the Si substrate, the rate was further increased (in steps, reaching up to 3.0-4.0) as well as the pulse energy (in steps, reaching up to 10-15 pJ) in order to collect a substantial amount of substrate atoms as quickly as possible before ending the run. The base temperature and base pressure within the APT chamber were maintained at 30 K and $3.2\times10^{-11}$ Torr, respectively.

**RESULTS AND DISCUSSION**

A representative cross-sectional transmission electron microscopy (TEM) image of an as-grown stack in Fig.1a shows the TL/ML/BL with a thickness of 220/160/60 nm, respectively, while dislocations are mainly visible close to the interface with the Ge-VS.[1,18] The reciprocal space mapping (RSM) recorded around the (224) diffraction peaks in Fig.1b indicates that the GeSn layers are compressively-strained. BL/ML diffraction peaks are close to the relaxation line (dashed line), revealing a low residual strain (-0.10% and -0.56%, respectively), whereas TL peak is vertically aligned with ML peak and thus is fully strained to the latter (-1.26% of strain).

Furthermore, the experimental lattice parameters are used to calculate the expected relaxed lattice parameter using the elastic constants and then retrieve the alloy composition following Vegard's law.[18] The Sn content in as-grown layers is 18/13/8at.% in the TL/ML/BL, respectively. The growth process described above yields high quality and atomically homogeneous GeSn materials leading to the demonstration of clear room temperature photoluminescence at wavelengths in the mid-infrared range.[1] However, few segregation "spots" occasionally appear on the $Ge_{0.82}Sn_{0.18}$ layer's mirror-like surface when the stacking thickness exceeds 400 nm. The Fig.1c (inset) shows an optical image of a particularly large spot, extending over 7 mm from its geometrical center (white dot) to its edge (red dot). The scanning electron micrograph (SEM) in Fig.1c reveals the presence of Sn droplets at the transition point between the mirror-like and segregated regions (red dot in Fig.1c, inset). Large Sn droplets with diameters of 5-15 µm (green arrow) propagate along the <110> direction leaving behind a straight trail of redeposited material (in light grey), similar to early observations.[12,19,20] These trails act as nucleation sites for smaller droplets (diameter well below 1 µm), which propagate along directions perpendicular to the main trail, leading to triangular areas formed by their paths (dashed lines in Fig.1c).

TEM images were acquired in the segregated area in order to investigate the material's microstructural properties right at the onset of phase separation. The cross-section shows multiple Sn droplets in the surface region, partially embedded in the GeSn layer (Fig.2a). A dense network of extended dislocations across the entire GeSn layer is visible, in striking contrast to the case of the mirror-like region (Fig.1a). The dashed lines indicates the interfaces positions in the GeSn stacking before phase separation. Additionally, the contrast change in the droplets suggests the presence of 2 regions with different composition and/or microstructure. Firstly, the energy-dispersive X-ray spectroscopy (EDX) map (Fig.2b) recorded around a large Sn droplet reveals the distribution of Sn in the cross-section. The unexpected ~40 nm-thick region at the droplet-layer interface exhibit a different Sn content from those expected for the two-equilibrium Sn- and Ge-rich phases. Indeed, a first estimate using EDX reveals a content of ~25at.% of Sn in this interfacial region, located above a much thicker depleted layer incorporating ~1at.% of Sn in average. This last value must be dealt with carefully since it is not known at this point if the material is still a homogeneous random alloy. The high-resolution TEM image and the corresponding fast Fourier transform (FFT) acquired on the Sn droplet, the defective Sn-rich layer, and the GeSn layer are shown in Fig.2c. A high number of stacking faults are visible in the interfacial layer. Yet, (002) and (220) diffraction spots positions in this defective layer and in the thicker layer underneath suggests that crystalline coherency is maintained at the interface.

To gain more insights into the structural properties of the segregation region, $2\theta$-$\omega$ X-ray diffraction (XRD) measurements around the (004) diffraction peak were performed on multiple positions across the segregated area (Fig.1c, inset), using an X-ray beam footprint of 1 mm×1.8 mm (Fig.3a). The center of the segregation spot is labelled as 0 mm (white dot in Fig.1c, inset) and the boundary with the mirror-like region is labelled as 7 mm (red dot in Fig.1c, inset). The idea behind this linear mapping is to probe the different regions of the spot to unravel the material bulk properties in the early and late stages of phase separation, supposedly located at the edges and at the centre of the spot, respectively. The reference XRD curve (black) was acquired on a mirror-like area of the sample. The Ge-VS peak is measured at 66.08°, while the peaks at lower angles are related to the GeSn layers. A first additional peak appears at 65.82° (violet arrow) and its intensity increases by a factor of ~3 when moving towards the center of the segregation spot, where a second high-intensity peak is also observed at 65.90° (red arrow). Based on RSM measurements (Figs.3b-c), the first and second peaks can be attributed to GeSn alloys with a Sn content of 1.5 and 1.1at.%, respectively. According to the peaks' positions relative to the relaxation line (white dashed line), these alloys are under a weak tensile strain, possibly due to the influence of the epitaxial interface with the Ge VS during Sn out-diffusion.

In the second APT dataset recorded for regions afar from the large droplets (Fig.5a), 4at.% Sn isoconcentration surfaces are added to the 3-D reconstruction to highlight the nonuniform distribution of Sn atoms. This nonuniformity is also visible in the compositional profile exhibited in Fig.5b (orange spheres). The Sn content rises up to ~10at.% in the upper and lower portions of the top layer, while a much lower content of 1-2at.% is observed elsewhere. Interestingly, Sn atoms seem to accumulate along vertical filaments. The radial concentration profiles, acquired along two orthogonal directions (green and light blue cylinders in Fig.5a) across one filament plotted in Fig.5c, reveal a nonuniform distribution of Sn varying between 0.7 and 3.2at.%. In both regions, the phase separation of GeSn is remarkably different from the expected transition towards the two thermodynamically stable phases incorporating either ~99.9at.% or ~1.0at.% of Sn.

Curiously enough, the Sn-enriched filaments revealed by APT (Fig.5a) are separated by approximately 100 nm and their morphology matches that of the dislocation lines identified in Fig.2a. Detailed and comparative APT and TEM studies also confirm that these filaments are Sn-decorated dislocations, reminiscent of Cottrell atmospheres.[21] Furthermore, the observed asymmetry in the radial Sn profile across each filament (Fig.5c) is similar to the behavior of solute atoms around a dislocation, where the tensile and compressive strained sides exhibit different solute concentrations.[21] These observations suggest that the formation of these dislocations is not only an efficient pathway to relax the epitaxial strain as the layer grow thicker, but it also triggers the phase separation by promoting Sn segregation toward the surface through pipe diffusion. The latter refers to an enhanced diffusivity in a short range around dislocations' core. The very few reports on this phenomenon in semiconductors suggested an increase in the diffusion coefficient by factor of ~$10^4$ .[22–24] This enhancement was also proposed to explain the diffusion at low temperature in defective materials such as cold-worked metals.[25–28] In the case of GeSn alloys, this process establish a clear relationship between the microstructure and material stability. It is noteworthy that, in the first stage of buffered growth, surface segregation by pipe diffusion is rather inhibited since dislocations are confined to the buffer layers (BL/ML in Fig.2a). This efficient defects engineering allows the TL to grow well above its theoretical critical thickness, estimated around 20 nm for a direct growth on the Ge VS.[29–32] However, when the total thickness of the GeSn stacking increases above 400 nm, the defective area extends and dislocations propagate toward the surface.[1,18] While the diffusion in the bulk is very slow,[33] the nucleation of threading dislocations and their propagation across the TL to reach the surface provide Sn atoms with a much higher mobility and a preferential path to diffuse and segregate on the surface.[23,34–36] It is noteworthy that similar atomistic and microstructural features are also observed when the phase separation is activated by annealing GeSn monocrystalline layers above the growth temperature. In recent annealing experiments realized on GeSn layers, the authors attributed the observed fast and brutal segregation of Sn in relaxed layers to the presence of dislocations, as opposed to diffusional mass transport in pseudomorphic layers.[37,38]

The transport of Sn from bulk to surface leads to the formation of self-propelled Sn-droplets. The motion of these droplets is likely to be a result of a gradient in surface Sn content, while the directionality of their guided motion can be a consequence of surface symmetry or atomic steps that are characteristics to the epitaxial growth of strained layers. Once the growth is interrupted and the substrate is cooled down, the droplets solidify and a defective interfacial layer seems to form as a result of this solidification. As the solubility of Ge in Sn decreases by ~2at.% when the sample is cooled down from the growth temperature to eutectic temperature (231°C), it is plausible that the defective layer underneath each solidified droplet (Fig.2a) forms from the expelled excess Ge from the cooling droplets.[13,14] This process yields a thin layer with a graded Sn content around 22at.% (Fig.4c). This relatively high Sn content might be dictated by the minimization of the mismatch strain at the droplet-GeSn interface during the solidification. Indeed, the preferential crystallographic orientation between the planes (-2-2 2) of GeSn and (301) of β-Sn (Fig.2c) seems to support this observation, in agreement with recent post-growth annealing studies.[39]

**CONCLUSIONS**

In summary, atomic-level insights into the complex behavior of Sn atoms at the onset of phase separation have been unraveled by combining advanced characterization techniques. In addition to the expected Sn surface segregation and the formation of regions at the Sn equilibrium composition of 1.0at.%, the diffusion of Sn in metastable GeSn also yields Sn-decorated filaments and bulk regions with a nonuniform Sn content in the range of ~1 to 11at.%. The latter are attributed to the formation and propagation of dislocations, which appears to facilitate the Sn transport toward the surface through pipe diffusion. 3-D atomic mapping indicates an asymmetric radial Sn profiles across these dislocation lines, in addition to the formation of a defective layer with a Sn content of ~22at.% at the interface between the Sn droplet and GeSn. Strain minimization between tetragonal Sn-rich and cubic Ge rich equilibrium phases has been suggested as a plausible reason for the formation of this layer. These studies highlight the key processes involved in the spontaneous phase separation during the out-of-equilibrium growth of strained layers and indicate the importance of dislocations in shaping the atomic pathways for solute diffusion and the associated formation of equilibrium phases. These observations are not only important to understand and optimize the growth of metastable layers but also to develop atomistic-level and predictive models to describe their behavior and stability.

**Notes**

The authors declare no competing financial interest.

## Acknowledgements

The authors thank J. Bouchard for the technical support with the CVD system. O.M. acknowledges support from NSERC Canada (Discovery, SPG, and CRD Grants), Canada Research Chairs, Canada Foundation for Innovation, Mitacs, PRIMA Québec, and Defence R&D Canada (Innovation for Defence Excellence and Security, IDEaS).

* corresponding author: oussama.moutanabbir@polymtl.ca

**Figure Captions**

**FIG. 1**. (a) Cross-sectional TEM image along zone axis [110] of as-grown GeSn TL/ML/BL/Ge-VS multilayer structure. The dashed lines highlight the interfaces between the GeSn layers; (b) RSM around the (224) Bragg reflection; (c) SEM image of the surface at the segregation spot edge. The green dashed lines indicate the smaller droplets' propagation front, at an angle of ~40° relative to the large droplet's trail. Inset: Photograph of a typical segregation spot.

**FIG. 2**. (a) Cross-sectional TEM image along zone axis [110] showing the segregated GeSn stacking and surface Sn droplets; (b) EDX map around a droplet embedded in the remaining GeSn layer; (c) HRTEM image revealing the interface between a Sn droplet, a thin defective GeSn layer and the segregated GeSn stacking, from top to bottom, respectively, with their corresponding indexed FFT patterns.

**FIG. 3**. (a) 2θ-ω XRD scans along the (004) reflection showing diffraction peaks from the Ge VS and low-Sn-content alloys; (b) RSM around the (224) Bragg reflection. Top: on the edge of segregation spot. Bottom: in the center of the segregation spot. The white dashed line represents the relaxation line.

**FIG. 4**. (a) SEM image of the surface in the segregation spot, approximately 3 mm from the center of the spot. A metal layer was locally deposited for the APT specimen preparation; (b) A typical 3-D reconstructed APT atom-by-atom map showing the interface between the bulk GeSn (lower part) and a droplet (upper part) in the region indicated by the violet arrow in Fig.4a; (c) Axial concentration profile of Sn across the 3D map in Fig.4b.

**FIG. 5**. (a) 3-D reconstructed APT atom-by-atom map from a region afar from droplets; (b) Axial concentration profile of Sn across the 3D map in Fig.5a. A profile from a non-segregated area is also shown for comparison; (c) Radial concentration profiles of Sn across the Sn-enriched line in Fig.5a, extracted from 15 nm-diameter cylinders.

# References

(1) Assali, S.; Nicolas, J.; Mukherjee, S.; Dijkstra, A.; Moutanabbir, O. Atomically Uniform Sn-Rich GeSn Semiconductors with 3.0–3.5 $\mu$m Room-Temperature Optical Emission. *Appl. Phys. Lett.* **2018**, *112* (25), 251903. https://doi.org/10.1063/1.5038644.

(2) Reboud, V.; Gassenq, A.; Pauc, N.; Aubin, J.; Milord, L.; Thai, Q. M.; Bertrand, M.; Guilloy, K.; Rouchon, D.; Rothman, J.; et al. Optically Pumped GeSn Micro-Disks with 16% Sn Lasing at 3.1 µm up to 180 K. *Appl. Phys. Lett.* **2017**, *111* (9), 092101. https://doi.org/10.1063/1.5000353.

(3) Margetis, J.; Yu, S.-Q.; Bhargava, N.; Li, B.; Du, W.; Tolle, J. Strain Engineering in Epitaxial $Ge_{1-x}Sn_x$: A Path towards Low-Defect and High Sn-Content Layers. *Semicond. Sci. Technol.* **2017**, *32* (12), 124006. https://doi.org/10.1088/1361-6641/aa7fc7.

(4) Senaratne, C. L.; Gallagher, J. D.; Aoki, T.; Kouvetakis, J.; Menéndez, J. Advances in Light Emission from Group-IV Alloys via Lattice Engineering and n-Type Doping Based on Custom-Designed Chemistries. *Chem. Mater.* **2014**, *26* (20), 6033–6041. https://doi.org/10.1021/cm502988y.

(5) Attiaoui, A.; Moutanabbir, O. Indirect-to-Direct Band Gap Transition in Relaxed and Strained $Ge_{1-x-y}Si_xSn_y$ Ternary Alloys. *J. Appl. Phys.* **2014**, *116* (6), 063712. https://doi.org/10.1063/1.4889926.

(6) Lan, H.-S.; Chang, S. T.; Liu, C. W. Semiconductor, Topological Semimetal, Indirect Semimetal, and Topological Dirac Semimetal Phases of $Ge_{1-x}Sn_x$ Alloys. *Phys. Rev. B* **2017**, *95* (20), 201201. https://doi.org/10.1103/PhysRevB.95.201201.

(7) Wirths, S.; Buca, D.; Mantl, S. Si–Ge–Sn Alloys: From Growth to Applications. *Prog. Cryst. Growth Charact. Mater.* **2016**, *62* (1), 1–39. https://doi.org/10.1016/J.PCRYSGROW.2015.11.001.

(8) Van Vechten, J. A.; Phillips, J. C. New Set of Tetrahedral Covalent Radii. *Phys. Rev. B* **1970**, *2* (6), 2160–2167. https://doi.org/10.1103/PhysRevB.2.2160.

(9) Loo, R.; Shimura, Y.; Ike, S.; Vohra, A.; Stoica, T.; Stange, D.; Buca, D.; Kohen, D.; Margetis, J.; Tolle, J. Epitaxial GeSn: Impact of Process Conditions on Material Quality. *Semicond. Sci. Technol.* **2018**, *33* (11), 114010. https://doi.org/10.1088/1361-6641/aae2f9.

(10) Weisshaupt, D.; Jahandar, P.; Colston, G.; Allred, P.; Schulze, J.; Myronov, M. Impact of Sn Segregation on $Ge_{1-x}Sn_x$ Epi-Layers Growth by RP-CVD. In *2017 40th International Convention on Information and Communication Technology, Electronics and Microelectronics (MIPRO)*; IEEE, 2017; pp 43–47. https://doi.org/10.23919/MIPRO.2017.7973388.

(11) Grant, P. C.; Dou, W.; Alharthi, B.; Grant, J. M.; Tran, H.; Abernathy, G.; Mosleh, A.; Du, W.; Li, B.; Mortazavi, M.; et al. UHV-CVD Growth of High Quality GeSn Using $SnCl_4$ : From Material Growth Development to Prototype Devices . *Opt. Mater. Express* **2019**, *9* (8), 3277. https://doi.org/10.1364/ome.9.003277.

(12) Tsukamoto, T.; Hirose, N.; Kasamatsu, A.; Mimura, T.; Matsui, T.; Suda, Y. Investigation of Sn Surface Segregation during GeSn Epitaxial Growth by Auger Electron Spectroscopy and Energy Dispersive X-Ray Spectroscopy. *Appl. Phys. Lett.* **2015**, *106* (5), 052103.

https://doi.org/10.1063/1.4907863.

(13) Olesinski, R. W.; Abbaschian, G. J. The Ge-Sn (Germanium-Tin) System. *Bull. Alloy Phase Diagrams* **1984**, *5* (3), 265–271. https://doi.org/10.1007/BF02868550.

(14) Feutelais, Y.; Legendre, B.; Fries, S. G. Thermodynamic Evaluation of the System Germanium — Tin. *Calphad* **1996**, *20* (1), 109–123. https://doi.org/10.1016/0364-5916(96)00018-1.

(15) Aubin, J.; Hartmann, J. M.; Gassenq, A.; Milord, L.; Pauc, N.; Reboud, V.; Calvo, V. Impact of Thickness on the Structural Properties of High Tin Content GeSn Layers. *J. Cryst. Growth* **2017**, *473*, 20–27. https://doi.org/10.1016/j.jcrysgro.2017.05.006.

(16) Aubin, J.; Hartmann, J. M.; Gassenq, A.; Rouviere, J. L.; Robin, E.; Delaye, V.; Cooper, D.; Mollard, N.; Reboud, V.; Calvo, V. Growth and Structural Properties of Step-Graded, High Sn Content GeSn Layers on Ge. *Semicond. Sci. Technol.* **2017**, *32* (9), 094006. https://doi.org/10.1088/1361-6641/aa8084.

(17) Piao, J.; Beresford, R.; Licata, T.; Wang, W. I.; Homma, H. Molecular-Beam Epitaxial Growth of Metastable $Ge_{1-x}Sn_x$ Alloys. *J. Vac. Sci. Technol. B Microelectron. Nanom. Struct.* **1990**, *8* (2), 221. https://doi.org/10.1116/1.584814.

(18) Assali, S.; Nicolas, J.; Moutanabbir, O. Enhanced Sn Incorporation in GeSn Epitaxial Semiconductors via Strain Relaxation. *J. Appl. Phys.* **2019**, *125* (2). https://doi.org/10.1063/1.5050273.

(19) Zhang, J.; Deng, X.; Swenson, D.; Hackney, S. A.; Krishnamurthy, M. Formation of Nanoscale Trenches and Wires as a Pathway to Phase-Separation in Strained Epitaxial Ge-Sn Alloys. *Thin Solid Films* **1999**, *357* (1), 85–89. https://doi.org/10.1016/S0040-6090(99)00481-2.

(20) Deng, X.; Yang, B.-K.; Hackney, S. A.; Krishnamurthy, M.; Williams, D. R. M. Formation of Self-Assembled Quantum Wires during Epitaxial Growth of Strained GeSn Alloys on Ge(100): Trench Excavation by Migrating Sn Islands. *Phys. Rev. Lett.* **1998**, *80* (5), 1022–1025. https://doi.org/10.1103/PhysRevLett.80.1022.

(21) Cottrell, A. H. Theory of Dislocations. *Prog. Met. Phys.* **1949**, *1* (C). https://doi.org/10.1016/0502-8205(49)90004-0.

(22) Legros, M.; Dehm, G.; Arzt, E.; Balk, T. J. Observation of Giant Diffusivity along Dislocation Cores. *Science* **2008**, *319* (5870), 1646–1649. https://doi.org/10.1126/science.1151771.

(23) Garbrecht, M.; Saha, B.; Schroeder, J. L.; Hultman, L.; Sands, T. D. Dislocation-Pipe Diffusion in Nitride Superlattices Observed in Direct Atomic Resolution. *Sci. Rep.* **2017**, *7* (1), 46092. https://doi.org/10.1038/srep46092.

(24) Volin, T. E.; Lie, K. H.; Balluffi, R. W. Measurement of Rapid Mass Transport along Individual Dislocations in Aluminum. *Acta Metall.* **1971**, *19* (4), 263–274. https://doi.org/10.1016/0001-6160(71)90092-7.

(25) Balliger, N. K.; Honeycombe, R. W. K. Coarsening of Vanadium Carbide, Carbonitride, and Nitride in Low-Alloy Steels. *Met. Sci.* **1980**, *14* (4), 121–133. https://doi.org/10.1179/030634580790426337.

(26) Gaboriaud, R. J. Dislocation Core and Pipe Diffusion in $Y_2O_3$. *J. Phys. D. Appl. Phys.* **2009**, *42* (13), 135410. https://doi.org/10.1088/0022-3727/42/13/135410.

(27) Watanabe, T.; Karashima, S. On the Strain-Enhanced Diffusion of α-Iron. *Phys. status solidi* **1970**, *42* (2), 749–756. https://doi.org/10.1002/pssb.19700420230.

(28) Varschavsky, A.; Donoso, E. Modelling the Kinetics of Solute Segregation to Partial Dislocations in Cold-Rolled Copper Alloys. *Mater. Lett.* **1997**, *31* (3–6), 239–245. https://doi.org/10.1016/S0167-577X(96)00273-X.

(29) People, R.; Bean, J. C. Calculation of Critical Layer Thickness versus Lattice Mismatch for $Ge_xSi_{1-x}$/Si Strained-layer Heterostructures. *Appl. Phys. Lett.* **1985**, *47* (3), 322–324. https://doi.org/10.1063/1.96206.

(30) People, R.; Bean, J. C. Erratum: Calculation of Critical Layer Thickness versus Lattice Mismatch for $Ge_xSi_{1-x}$/Si Strained-layer Heterostructures [Appl. Phys. Lett. **4 7**, 322 (1985)]. *Appl. Phys. Lett.* **1986**, *49* (4), 229–229. https://doi.org/10.1063/1.97637.

(31) Wang, W.; Zhou, Q.; Dong, Y.; Tok, E. S.; Yeo, Y.-C. Critical Thickness for Strain Relaxation of $Ge_{1-x}Sn_x$ ($x \leq 0.17$) Grown by Molecular Beam Epitaxy on Ge(001). *Appl. Phys. Lett.* **2015**, *106* (23), 232106. https://doi.org/10.1063/1.4922529.

(32) Gencarelli, F.; Vincent, B.; Demeulemeester, J.; Vantomme, A.; Moussa, A.; Franquet, A.; Kumar, A.; Bender, H.; Meersschaut, J.; Vandervorst, W.; et al. Crystalline Properties and Strain Relaxation Mechanism of CVD Grown GeSn. *ECS J. Solid State Sci. Technol.* **2013**, *2* (4), P134–P137. https://doi.org/10.1149/2.011304jss.

(33) Wang, W.; Dong, Y.; Zhou, Q.; Tok, E. S.; Yeo, Y.-C. Germanium–Tin Interdiffusion in Strained Ge/GeSn Multiple-Quantum-Well Structure. *J. Phys. D. Appl. Phys.* **2016**, *49* (22), 225102. https://doi.org/10.1088/0022-3727/49/22/225102.

(34) Cottrell, A. H.; Bilby, B. A. Dislocation Theory of Yielding and Strain Ageing of Iron. *Proc. Phys. Soc. Sect. A* **1949**, *62* (1), 49–62. https://doi.org/10.1088/0370-1298/62/1/308.

(35) Bonef, B.; Shah, R. D.; Mukherjee, K. Fast Diffusion and Segregation along Threading Dislocations in Semiconductor Heterostructures. *Nano Lett.* **2019**, *19* (3), 1428–1436. https://doi.org/10.1021/acs.nanolett.8b03734.

(36) Dobrokhotov, E.́ V. Diffusion in Dislocation Germanium and the Model of a Liquid Dislocation Core. *Phys. Solid State* **2005**, *47* (12), 2257–2261. https://doi.org/10.1134/1.2142887.

(37) Zaumseil, P.; Hou, Y.; Schubert, M. A.; von den Driesch, N.; Stange, D.; Rainko, D.; Virgilio, M.; Buca, D.; Capellini, G. The Thermal Stability of Epitaxial GeSn Layers. *APL Mater.* **2018**, *6* (7), 076108. https://doi.org/10.1063/1.5036728.

(38) Driesch, N. von den; Wirths, S.; Troitsch, R.; Mussler, G.; Breuer, U.; Moutanabbir, O.; Grützmacher, D.; Buca, D. Thermally Activated Diffusion and Lattice Relaxation in (Si)GeSn Materials. **2019**.

(39) Groiss, H.; Glaser, M.; Schatzl, M.; Brehm, M.; Gerthsen, D.; Roth, D.; Bauer, P.; Schäffler, F. Free-Running Sn Precipitates: An Efficient Phase Separation Mechanism for Metastable $Ge_{1-x}Sn_x$ Epilayers. *Sci. Rep.* **2017**, *7* (1), 16114. https://doi.org/10.1038/s41598-017-16356-8.

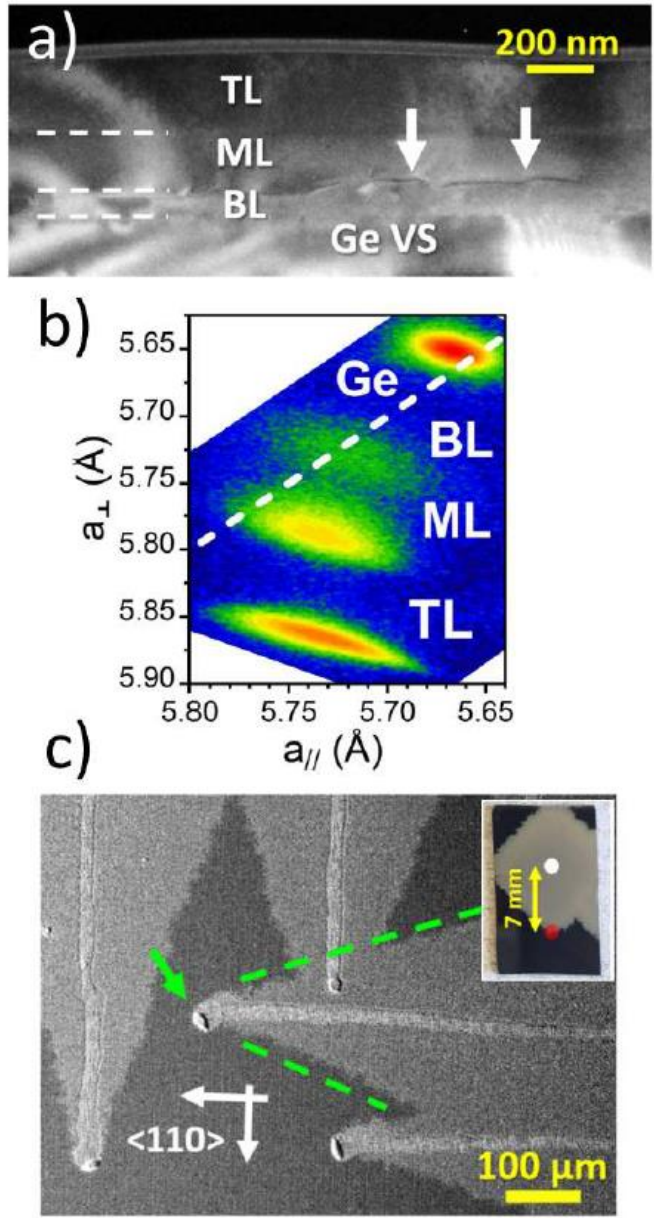

**Figure 1**

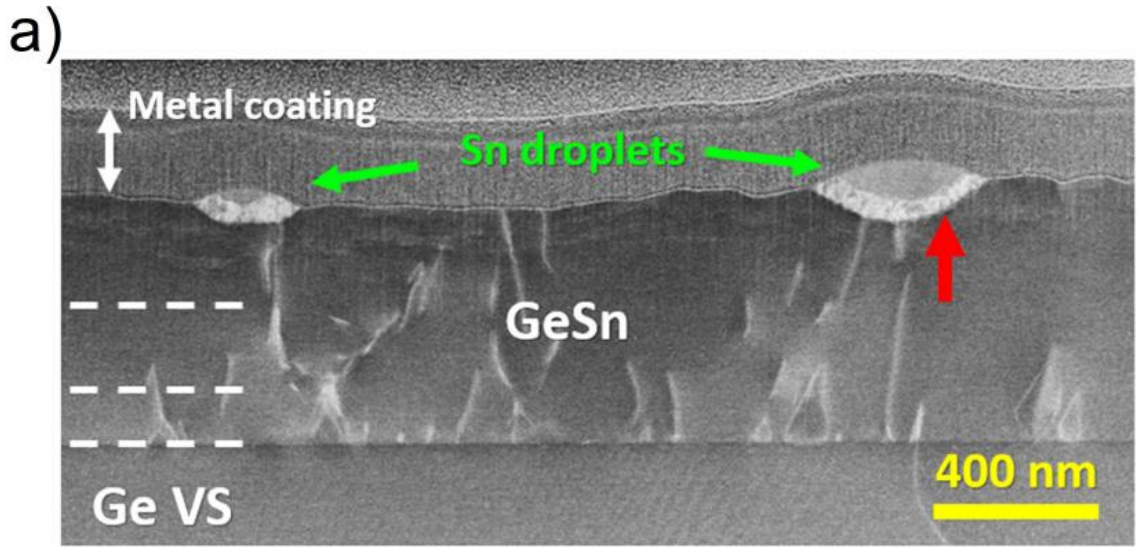
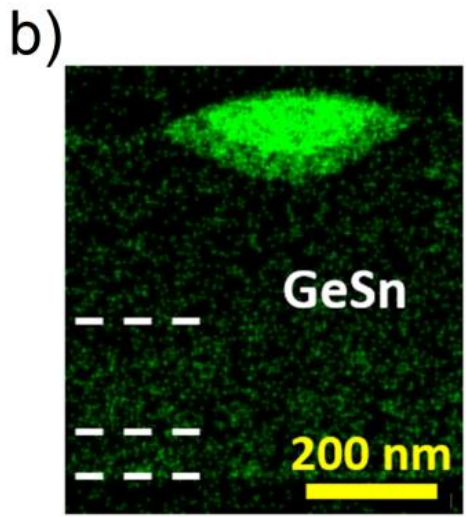
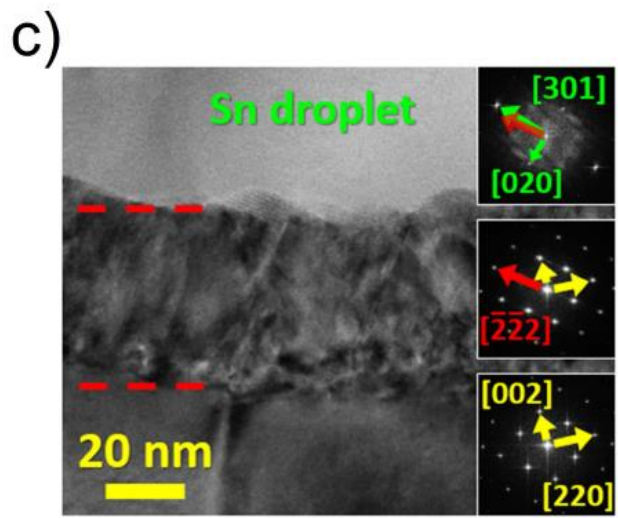

**Figure 2**

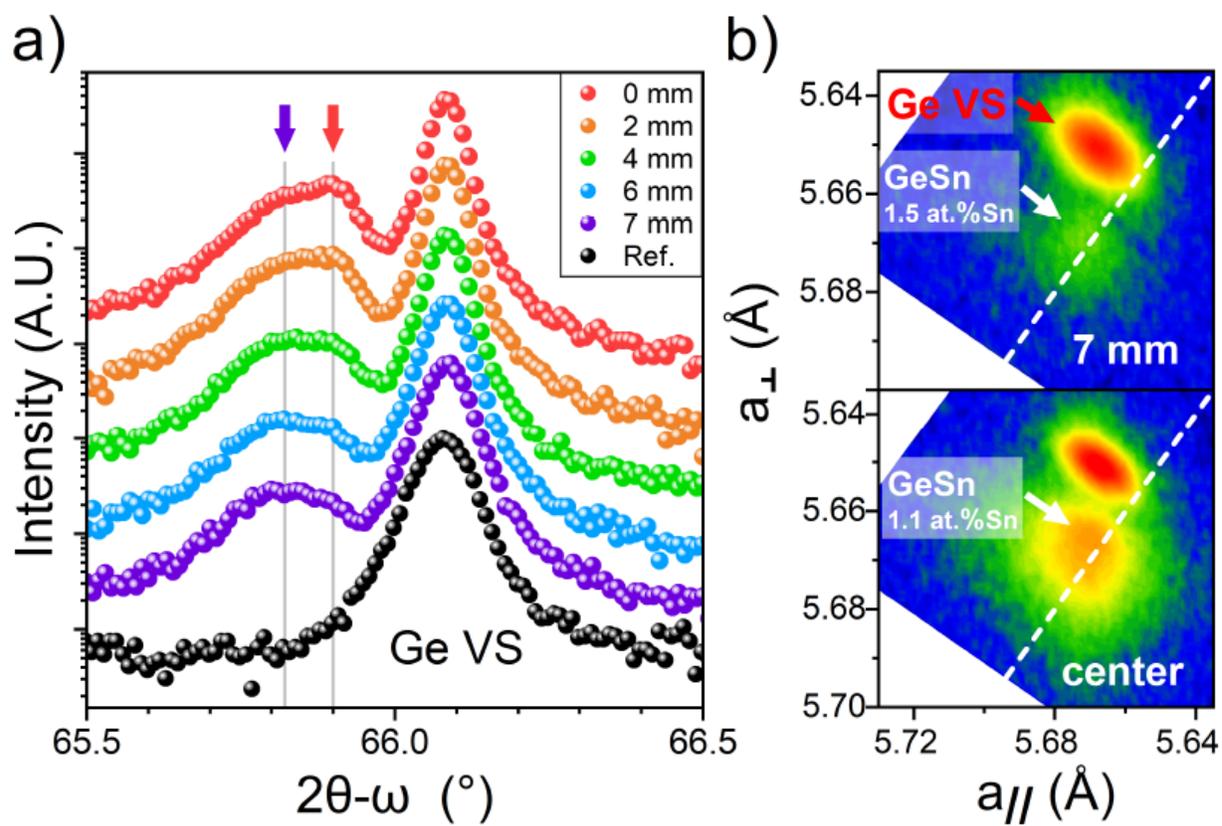

**Figure 3**

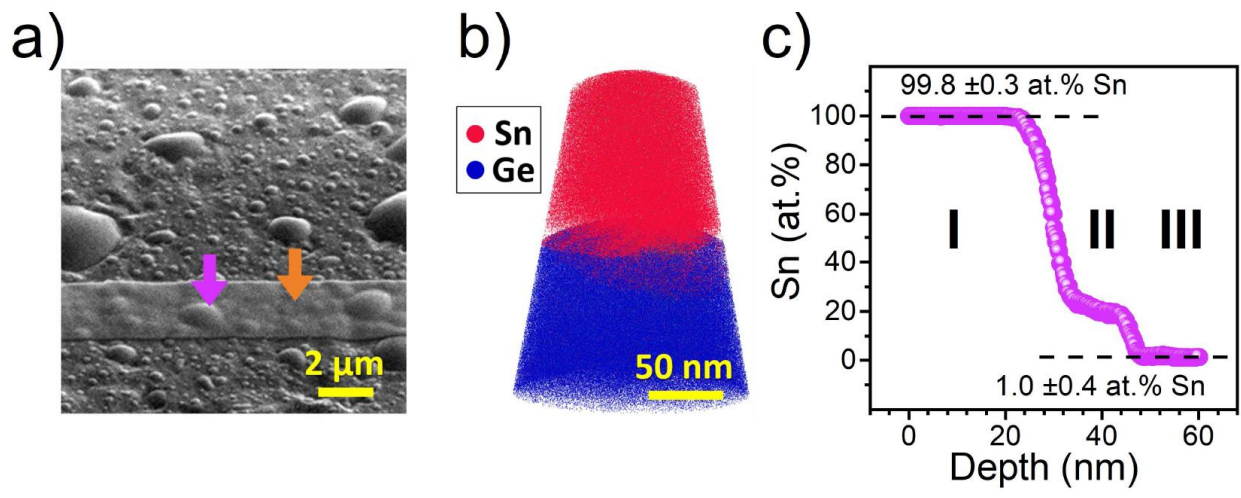

**Figure 4**

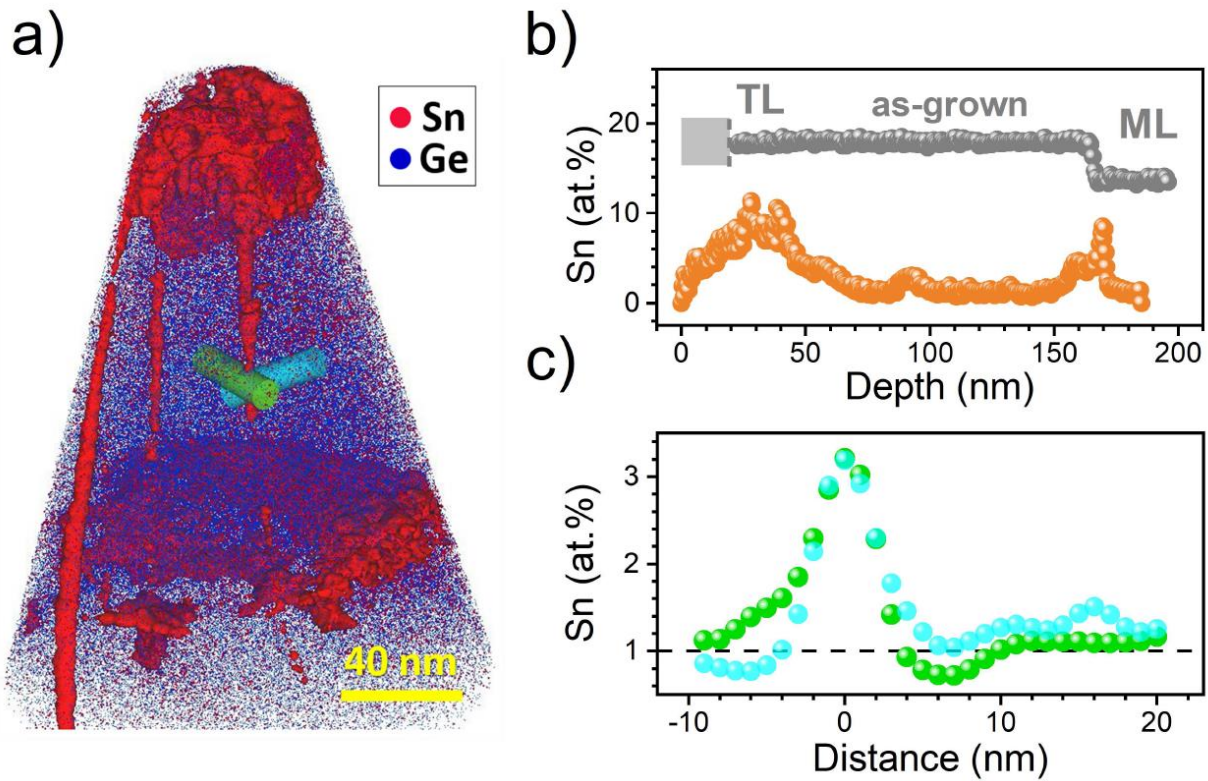

**Figure 5**